\documentclass[amsmath,amssymb,pra,twocolumn]{revtex4}
\usepackage[cp1251]{inputenc}
\usepackage[english]{babel}
\usepackage{color}
\usepackage{graphicx}
\usepackage{dcolumn}
\usepackage{braket}
\usepackage{amsmath}
\usepackage{natbib}

\numberwithin{equation}{section}
\begin{document}

%\bibliographystyle{unsrt}
%\bibliography{references}

\title{Temporal interference effects in noncollinear and frequency-nondegenerate spontaneous parametric down-conversion}

\author{M. V. Fedorov$^{1,2}$}
\author{A. A. Sysoeva$^{1,3}$, S. V. Vintskevich$^{1,3}$, D. A. Grigoriev$^{1,3}$}
\address{$^1$A.M.~Prokhorov General Physics Institute,
 Russian Academy of Sciences, 38 Vavilov st., Moscow, 119991, Russia}
 \address{$^2$National Research University Higher School of Economics, 20 Myasnitskaya Ulitsa,
Moscow, 101000, Russia}
\address{$^3$Moscow Institute of Physics and Technology, Dolgoprudny, Moscow Region, Russia}

\date{\today}

\begin{abstract}
We consider regimes of Spontaneous Parametric Down-Conversion both noncollinear and nondegenerate in frequencies. Parameters characterizing degrees of noncollinearity and of nondgeneracy are defined, and they are shown to be not independent of each other. At a given degree of nondegeneracy the emitted photons are shown to propagate along two different cones opening angles of which are determined by the degree of nondegeneracy. Based on this, the degree of nondegeneracy can be controlled by means of the angular selection of photons, e.g., with the help of appropriately installed slits. For such selected photons their wave functions are found depending on two frequency or on two temporal variables. Interference effects arising in such states are tested by analyzes of the Hong-Ou-Mandel type scheme  with the varying delay time in one of two channels and with photons from two channels sent to the beamsplitter. The temporal pictures arising after the beamsplitter are found to demonstrate extremely strong interference exhibiting itself in formation of finite-size temporal combs filled with quantum beats. Parameters of combs depend on the degree of nondegeneracy, and the physical reasons of this dependence are clarified.
\end{abstract}

\maketitle

\section{Introduction}
Spontaneous Parametric Down-Conversion (SPDC) is the effect which is well known (since 1967 \cite{Klyshko,Harris,Magde,Kleinman,Burnham}) and widely studied. Nowadays, SPDC sources are widely used throughout the world, and they can be considered as the main tool of numerous experiments in the fields of quantum optics and quantum information. Regimes of SPDC are rather well known also. They can be differentiated by the type of phase matching (I or II), by collinear or noncollinear propagation of emitted photons, by observation of emitted photons with coinciding or different frequencies (frequency-degenerate or nondegenrate processes), etc. In this work we consider a general case when the type-I SPDC process is both noncollinear and frequency-nondegenerate. In this formulation the degrees of noncollinearity and frequency-nondegeneracy are not independent of each other. Rather simple formulas are obtained describing explicitly connection between the defined below  below parameters of nondegeneracy and of noncollinearity. Existence of this connection can be used for controlling the degree of frequency-nondegeneracy by means of angular selection selection of photons to be registered. Such procedure can be realized with the help of appropriately installed two or four slits. In these schemes we find two-frequency biphoton wave function, parameters of which are functions of the degree of nondegeneracy. The double Fourier transformation is used to get the wave function depending on two temporal variables which are interpreted as the arrival times of photons to a detector or to a beamsplitter. A scheme with the beamsplitter is used for analyzing coherent features and interference phenomena of the arising states in the frame of the Hong-Ou-Madel (HOM) effect \cite{HOM}. For four-slit scheme this effect is found to have a rather peculiar form, with many oscillations of the coincidence probability in dependence on the delay time in one of two channels and with formation  of finite-size temporal-comb structures. Various aspects of similar analyzes were considered in a series of works \cite{Rubin,Shi-Serg,Sergienko,ASL,Abouraddy,Beduini,Paulina,AILv,Guo,Agata}. Whenever it's reasonable, the results of the present work will be compared with those obtained earlier by us or other authors.

Note that in addition to fundamental interest to physics of phenomena arising in frequency-nondegenerate regimes of SPDC, they can be important also for applications such as, e. g., IR spectroscopy, because in the extreme cases of very high degree of nondegeneracy the longer-wavelength emitted photons can reach the IR diapason \cite{GKhK,SPK}. This new direction of investigations is an additional motivation for performing the presented below general analysis of the frequency-nondegenerate noncollinear regimes of SPDC.

\section{Nondegeneracy of central frequencies}

In this work we consider only the type-I phase matching which means that the pump propagates in a crystal as an extraordinary wave and some of its photons decay for two ordinary-wave SPDC photons, $e\rightarrow o+o$. The wave function characterizing angular and frequency distributions of emitted photons is well known to have the form
\begin{equation}
 \label{w-f-gen}
 \Psi\propto E_p\, {\rm sinc}(L\Delta/2),
\end{equation}
where $E_p$ and $\Delta$ are the pump field-strength amplitude and the phase mismatch, ${\rm sinc}(x)=\sin x/x$, and $L$ is the length of a crystal along the pump propagation axis $0z$. In a general case both $E_p$ and $\Delta$ depend on angular and frequency variables of two emitted photons.   Let us assume that spectra of the pump and of emitted photons are relatively narrow and concentrated around the corresponding central frequencies, $\omega_p^{(c)}\equiv\omega_0$ and  $\omega_1^{(c)}=\omega_h$ (high) and $\omega_2^{(c)}=\omega_l$ (low). In a general case the central frequencies of emitted photons $\omega_h$ and $\omega_l$ can be different from each other though their sum is assumed to be equal to the pump central frequency, $\omega_h+\omega_l=\omega_0$ which corresponds to the energy conservation rule. The case $\omega_h\neq\omega_l$ corresponds to the SPDC process nondegenerate with respect to the central frequencies of emitted photons.  The degree of nondegeneracy can be characterized by a dimensionless parameter $\xi$ ($0\leq\xi<1$)
\begin{equation}
 \label{ksi}
 \xi=\frac{\omega_h-\omega_l}{\omega_0},
\end{equation}
in terms of which
\begin{equation}
 \label{omega hl via xi}
 \omega_{h,l}=\omega_0\frac{1\pm\xi}{2}\quad{\rm and}\quad \lambda_\pm^{(c)}=\frac{2\pi c}{\omega_{h,l}}=\frac{2\lambda_p^{(c)}}{1\pm\xi},
\end{equation}
where $\lambda_+^{(c)}$ and $\lambda_-^{(c)}$ are the central wavelengths of higher- and lower-frequency emitted photons in dependence of their spectra on wavelengths rather than frequencies.
\section{Phase matching}
\begin{figure}[h]
\centering\includegraphics[width=8 cm]{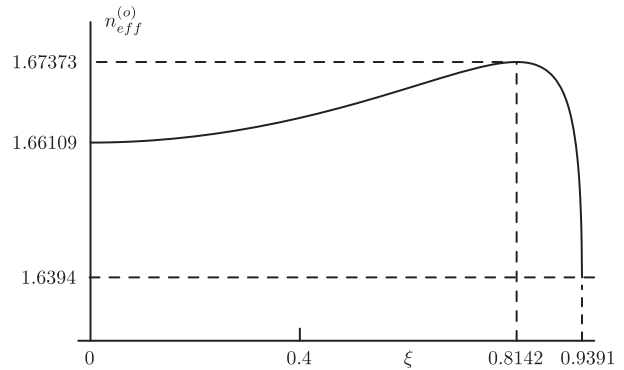}
\caption{{\protect\footnotesize {The effective ordinary-wave refractive index  $n_{\rm eff}^{(o)}$ as a function of the nondegeneracy parameter $\xi$}}}\label{Figl}
\end{figure}

\begin{figure}[h]
\centering\includegraphics[width=8 cm]{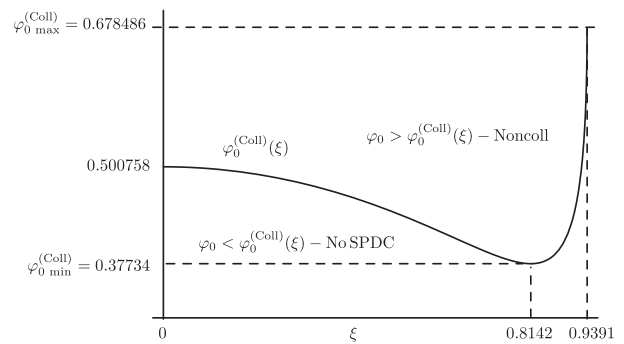}
\caption{{\protect\footnotesize {The angle $\varphi_0^{(\rm Coll)}(\xi)$ between the crystal optical axis and the pump-propagation direction $0z$ at which the SPDC process is collinear but frequency-nondegenerate.}}}\label{Fig2}
\end{figure}

Let us consider first the collinear frequency-nondegenerate case with frequencies of emitted photons equal exactly to $\omega_h$ and $\omega_l$. Then the phase mismatch is given by
\begin{equation}
 \label{mismatch}
 \Delta_0=k_p-k_1-k_2=\frac{2\pi}{\lambda_p}\left(n_p(\varphi_0)-n_{\rm eff}^{(o)}(\xi)\right),
\end{equation}
where $k_p$, $k_1$, and $k_2$ are absolute values of the pump- and emitted-photon wave vectors in a crystal; $n_p(\varphi_0)$ is the refractive index of the pump for its propagation strictly along the $z$-axis; $\varphi_0$ is the angle between the crystal optical axis and the axis $0z$. The effective ordinary-wave refractive index $n_{\rm eff}^{(o)}(\xi)$ is introduced in Eq. (\ref{mismatch}) to reduce it to the form similar to that occurring in the frequency-degenerate case,  $\Delta_0^{\rm deg}=(2\pi/\lambda_p)(n_p-n_o)$ with $n_o=n_o(2\lambda_p)$ being the ordinary-wave refractive index. In the nondegenerate case the effective ordinary-wave refractive index is given by
\begin{equation}
 \label{n-eff}
 n_{\rm eff}^{(o)}(\xi)=\frac{1+\xi}{2}n_o\left(\frac{2\lambda_p}{1+\xi}\right)+
 \frac{1-\xi}{2}n_o\left(\frac{2\lambda_p}{1-\xi}\right).
\end{equation}
Here and below we make all estimates for a BBO crystal of the length $L=0.5\,{\rm cm}$ and the pump wavelength $\lambda_p=0.4047\,\mu{\rm m}$. For these parameters the dependence  $n_{\rm eff}^{(o)}(\xi)$ is shown in Fig. \ref{Figl}.
Already from this picture we find that the maximal value of the nondegeneracy parameter $\xi$ leaving emitted photons in the transparency window of the crystal BBO equals $\xi_{\max}=0.9391$, and this value corresponds to the maximal achievable wavelength of the lower-frequency photon $\lambda_{-\,\max}=13.29\,\mu{\rm m}$.
Collinear frequency-nondegenerate regime occurs when the phase mismatch $\Delta_0$ (\ref{mismatch}) turns zero, or when $n_p(\varphi_0)=n_{\rm eff}^{(o)}(\xi)$. Solution of this equation, $\varphi_0^{(\rm Coll)}(\xi)$, is shown in Fig.  \ref{Fig2}.

The curve in Fig. \ref{Fig2} reflects all features of the effective refractive index shown in Fig. \ref{Figl}. In particular, this curve confirms once again that the maximal achievable values of the degree of nondegeneracy $\xi_{\max}$ and of the wavelength $\lambda_{-\,\max}$ of the emitted higher-frequency photon are equal to 0.9391 and $13.29\,\mu{\rm m}$, respectively. Also it's worth noting that the curves $n_{\rm eff}(\xi)$ and $\varphi_0^{\rm Coll}(\xi)$ have coinciding positions of their maximum and minimum, respectively, at $\xi= 0.8142$. In fact, as we'll see below, this is a very special point important not only for these curves but also for spectral features of noncollinear nondegenerate biphoton states.

All points at the curve in Fig. \ref{Fig2} correspond to pairs of parameters $(\varphi_0,\xi)$ at which SPDC is collinear. Thus, compared to the case of frequency-degenerate SPDC, transition to the frequency-nondegenerate regimes extends significantly the existence conditions of the collinear SPDC: the collinear regime can be realized at any orientations of the crystal optical axis in the whole interval between $\varphi_{0\,\min}^{(\rm Coll)}=0.37734$ to $\varphi_{0\,\max}^{(\rm Coll)}=0.678486$, if only the degree of nondegeneracy is appropriately chosen.
All points above the curve in Fig. \ref{Fig2} correspond to $n_p(\varphi_0)-n_{\rm eff}^{(o)}(\xi)<0$, and this is the region of nocollinear SPDC. And all points below the curve in Fig. \ref{Fig2} correspond to $n_p(\varphi_0)-n_{\rm eff}^{(o)}(\xi)>0$ when SPDC does not exist.

\section{Noncollinear frequency-nondegenerate regimes}
If $\Delta_0=\frac{2\pi}{\lambda_p}(n_p(\varphi_0)-n_{\rm eff}^{(o)}(\xi))<0$, this negative term can be compensated by a positive term determined by the first-order expansion of the phase mismatch in squared transverse components of wave vectors of emitted photons
\begin{equation}
 \label{k-perp-squared}
 \Delta_1^{\perp}=\frac{({\vec k}_{1\,\perp}-{\vec k}_{2\,\perp})^2}{8}\left(\frac{1}{k_1}+\frac{1}{k_2}\right).
\end{equation}
In any given plane $(x,z)$,
\begin{equation}
 \label{theta}
  k_{1,2\,x}=\frac{\pi}{\lambda_p}(1\pm\xi)\theta_{1,2\,x}
  \equiv\frac{\pi}{\lambda_p}{\widetilde\theta}_{1,2\,x},
\end{equation}
where $\theta_{1,2\,x}$ are photon propagation angles with respect to $0z$-axis in a free space after the crystal, and the notation   ${\widetilde\theta}_{1,2\,x}=(1\pm\xi)\,\theta_{1,2\,x}$ is used instead of $\theta_{1,2\,x}$ for shortening intermediate formulas. In contrast to  $k_{1,2\,x}$, the terms $1/k_1$ and $1/k_2$ in Eq. (\ref{k-perp-squared}) are absolute values of photon wave vectors in the crystal, with appropriate refractive indices taken into account, which gives
\begin{equation}
 \label{k-perp-squared-via theta}
 \frac{1}{k_1}+\frac{1}{k_2}=\frac{2\lambda_p}{\pi N_{\rm eff}(\xi)}\;{\rm and}\;
 \Delta_1^{\perp}=\frac{\pi}{4\lambda_p}\frac{({\widetilde\theta}_{1\,x}
 -{\widetilde\theta}_{2\,x})^2}{N_{\rm eff}(\xi)},
\end{equation}
where the function $N_{\rm eff}(\xi)$,
\begin{equation}
 \label{N-eff}
 N_{\rm eff}(\xi)=\frac{
 (1-\xi^2)n_o\left(\frac{2\lambda_p}{1+\xi}\right)n_o\left(\frac{2\lambda_p}{1-\xi}\right)}
 {n_{\rm eff}(\xi)},
\end{equation}
Note that at $\xi\neq 0$ $N_{\rm eff}(\xi)\neq n_{\rm eff}(\xi)$ though $N_{\rm eff}(0)= n_{\rm eff}(0)=n_o(2\lambda_p)$.

The term  $\Delta_0$ (\ref{mismatch}) in the phase mismatch can be redenoted as
\begin{equation}
  \Delta_0=\frac{2\pi}{\lambda_p}(n_p(\varphi_0)-n_{\rm eff}^{(o)}(\xi))=
   \label{Delta0 via theta-0}
   -\frac{\pi}{\lambda_p}\frac{\theta_0^2(\xi,\varphi_0)}{N_{\rm eff}(\xi)}.
\end{equation}
Eq. (\ref{Delta0 via theta-0}) is a definition of the function $\theta_0^2(\xi,\varphi_0)$ explicitly given by:
\begin{equation}
 \label{theta-0}
 \theta_0^2(\xi,\varphi_0)=2N_{\rm eff}(\xi)\left(n_{\rm eff}^{(o)}(\xi)-n_p(\varphi_0)\right).
\end{equation}
This expression is analogous to that occurring in the case of frequency-degenerate SPDC \cite{PRA-18}, $\theta_0=2n_o(n_0-n_p)$. Generalization for the nondegenerate case consists in the replacement of the factor $n_0$ in front of the difference $(n_0-n_p)$ by $N_{\rm eff}(\xi)$ and the term $n_o$ in this  difference by $n_{\rm eff}(\xi)$. Below differences in interpretation of the function $\theta_0(\xi)$ in the cases $\xi=0$ and $\xi \neq 0$ will be discussed in more details.

Combined together, Eqs. (\ref{k-perp-squared-via theta}) and (\ref{Delta0 via theta-0}) give the following expression for the sinc-function part of the angular biphoton wave function (\ref{w-f-gen})
\begin{gather}
 \nonumber
 \Psi_{\rm sinc}\propto{\rm sinc} \left[\frac{{(\widetilde\theta}_{1\,x}-{\widetilde\theta}_{2\,x})^2-4\theta_0^2}
 {8\theta_0(\delta\theta)_L}\right]\approx\\
 \label{2-sinc}
 {\rm sinc}\left[\frac{{\widetilde\theta}_{1\,x}-{\widetilde\theta}_{2\,x}-2\theta_0}
 {2(\delta\theta)_L}\right]+{\rm sinc}\left[\frac{{\widetilde\theta}_{1\,x}-
 {\widetilde\theta}_{2\,x}+2\theta_0}
 {2(\delta\theta)_L}\right],
\end{gather}
where
\begin{equation}
 \label{Delta theta L}
 (\delta\theta)_L=\frac{\lambda_p}{\pi L}\frac{N_{\rm eff}}{\theta_0}
 =\frac{\lambda_p}{2\pi L}\frac{\theta_0}{n_{\rm eff}^{(o)}-n_p}.
\end{equation}
In the second line of Eq. (\ref{2-sinc}) the single sinc-function with the argument quadratic in $({\widetilde\theta}_{1\,x}-{\widetilde\theta}_{2\,x})$ is replaced by the sum of two sinc-functions with arguments linear in the sum and difference of angles ${\widetilde\theta}_{1\,x}-{\widetilde\theta}_{2\,x}\pm 2\theta_0$. This is a rather usual approximation \cite{Scr} valid under the condition of sufficiently degree of noncollinearity, $\theta_0\gg (\delta\theta)_L$, or
\begin{equation}
 \label{2 sinc cond}
 n_{\rm eff}^{(o)}-n_p\gg\frac{\lambda_p}{2\pi L}\sim 10^{-4},
\end{equation}
which is easily satisfied.

The sinc-parts of the wave function (\ref{2-sinc}) have to be multiplied by the angular part of the pump field strength $E_p$
\begin{equation}
 \label{pump-angle}
 E_p^{\rm angle}\propto\exp\left[-\frac{\pi^2w^2}{2\lambda_p^2}
 ({\widetilde\theta}_{1\,x}+{\widetilde\theta}_{2\,x})^2\right],
\end{equation}
where $w$ is the pump waist.
Both factors together determine central values of the angles ${\widetilde\theta}_{1\,x}$ and ${\widetilde\theta}_{2\,x}$:
${\widetilde\theta}_{1\,x}=\pm\theta_0$ and ${\widetilde\theta}_{2\,x}=\mp\theta_0$. But ${\widetilde\theta}_{1,2\,x}$ are not yet the true propagation directions of emitted photons. In accordance with Eq. (\ref{theta}) the true propagation directions of emitted photons are given by
$\theta_{1\,x}=\theta_0/(1+\xi)$ and $\theta_{2\,x}=\theta_0/(1-\xi)$ or
$\theta_{1\,x}=\theta_0/(1-\xi)$  and $\theta_{2\,x}=\theta_0/(1+\xi)$.

\begin{figure}[t]
\centering\includegraphics[width=5 cm]{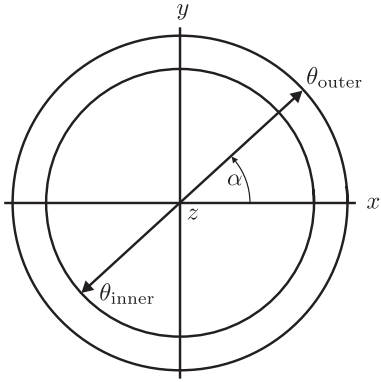}
\caption{{\protect\footnotesize {Inner and outer cones and their opening angles (\ref{inner-outer}).Arrows indicate location of photons in a given pair. Azimuthal angle $\alpha$ can take any values from 0 to $2\pi.$ }}}\label{Fig3}
\end{figure}

In fact, as the plane $(x,z)$ is arbitrary, these equations determine two propagation cones of photons, the outer and inner ones, with the cone axes coinciding with pump propagation direction $0z$ and with the cone opening angles equal to
\begin{equation}
 \label{inner-outer}
 \theta_{\rm outer}\equiv\theta_-=\frac{\theta_0}{1-\xi}\; {\rm and}\; \theta_{\rm inner}\equiv\theta_+=\frac{\theta_0}{1+\xi}.
\end{equation}
In all cases the lower-frequency photons propagate along the outer cone and the higher-frequency photons - along the inner cone, and in each pair of photons their propagation directions belong to opposite ends of diameters of the cone section by a plane perpendicular to the $z$-axis (Fig.\ref{Fig3}).

Though the function $\theta_0\equiv\theta_0(\xi,\varphi_0)$ itself is not an opening angle of any cones for photon propagation, it can be considered as the parameter characterizing the degree of noncollinearity, as well as $\xi$ (\ref{ksi}) is the nondegeneracy parameter. Then Eq. (\ref{theta-0}) [together with Eqs. (\ref{n-eff}) and (\ref{N-eff})] can be considered as the equation establishing connection between the degrees of noncollinearity and nondegeneracy.

\begin{figure}[h]
\centering\includegraphics[width=8 cm]{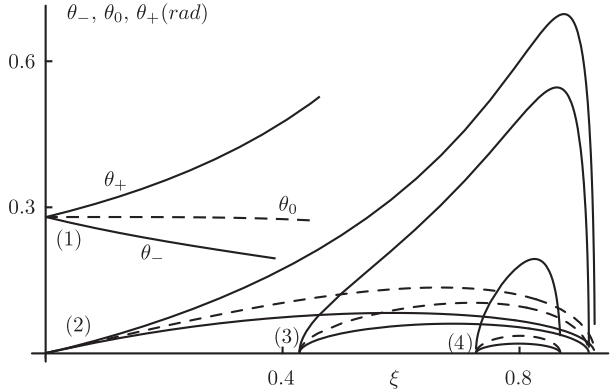}
\caption{{\protect\footnotesize {The functions $\theta_-(\xi),\theta_+(\xi)$,} and $\theta_0(\xi)$ (dashed lines) at $\varphi_0=0.7\,(1),\,0.5007589\,(2), 0.46\,(3)$ and $0.39\,(4).$   }}\label{Fig4}
\end{figure}

In Fig. \ref{Fig4} the angles $\theta_0$, $\theta_+$ and $\theta_-$ are shown  as functions of the nondegeneracy parameter $\xi$ at a series of values of the angle $\varphi_0$ between the crystal optical axis and the pump propagation direction $0z$.

Thus, for any given $\xi$ and $\varphi_0$ the central frequencies and cone-opening angles of photons arise in pars, $(\omega_h,\theta_{\rm inner})$ and $(\omega_l,\theta_{\rm outer})$. At values of the crystal-orientation angle $\varphi_0$ not too close to $\varphi_{0\,\min}=0.37734$ (see Fig. \ref{Fig2}), both the cone-opening angles $\theta_{\pm}(\xi)$ and central frequencies of emitted photons $\omega_{h,l}(\xi)$ vary continuously with varying nondegeneracy parameter $\xi$, if its value is not controlled at all. This means that a general picture of emission from a crystal is multicolored and multidirectional, like in a rainbow. In principle, as central frequencies of emitted photons depend on $\xi$, one can select photons with a given value of the nondegeneracy parameter by installing at the exit from the crystal a spectral filter with two transparency windows, around $\omega_h=\omega_0\frac{1+\xi}{2}$ and $\omega_l=\omega_0-\omega_h=\omega_0\frac{1-\xi}{2}$. This will provide automatically the angular selection of photons propagating only along two cones with the opening angles $\theta_{\rm inner}(\xi)$ and $\theta_{\rm outer}(\xi)$ (\ref{inner-outer}) and with $\xi=\frac{1}{2}(\omega_h-\omega_l)$.

\begin{figure}[h]
\centering\includegraphics[width=8 cm]{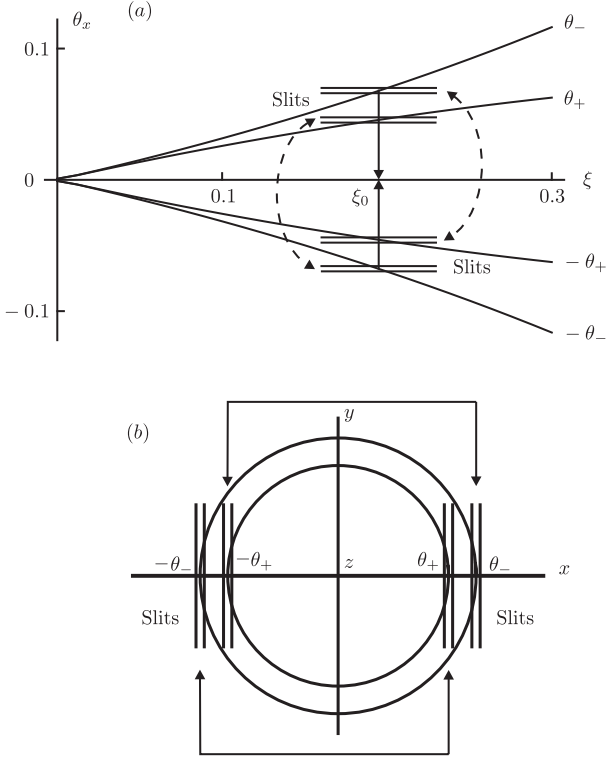}
\caption{{\protect\footnotesize {$(a)$ Cone opening angles at $\varphi_0=0.5007589$ and positions of slits photons providing a given value of the nondegeneracy parameter $\xi_0=0.2$. Dashed lines with arrows indicate pairs of slits through which photons of all given SPDC pairs propagate. $(b)$ The same in terms of real cones.; adjoint slits for photons of given pairs are indicated by brackets with arrows. }}}\label{Fig5}
\end{figure}

Another way of getting the same result is related to angular selection of photons.
As both the cone-opening angles $\theta_-$ and $\theta_+$ and the difference between them depend in on the nondegeneracy parameter $\xi$, angular selection can be realized with the help of slits. An example of a possible slit-installation scheme is shown in the diagram of  Fig. \ref{Fig5}. The picture \ref{Fig5}$(a)$ the angle between the crystal optical axis and the $z$-axis is taken equal to $\varphi_0=0.5007589$. In this example the SPDC process is collinear at $\xi=0$ and noncollinear at $\xi>0$. The slits are shown installed in positions appropriate for selection of photons with the nondegeneracy parameter $\xi_0=0.2$. The picture \ref{Fig5}$(b)$ represents the same $4-{\rm slit}$ scheme of measurements in terms of real cones.

Spectral features of biphoton states formed by means  angular selection will be discussed in sections 5 and 6. As a concluding remark for this section, note that in fact, the slit selection mentioned here and below can mean also selection by means of other devices, e.g., by appropriately installed fibers to be used for controlled transportation of photons to detectors, or beamsplitters, etc.

\section{Angular-frequency and temporal biphoton wave functions}
Until now it was assumed that photon frequencies coincide exactly either with $\omega_h$ or $\omega_l$. Let us find now corrections to this approximation, i.e., let us find corrections linear in deviations from central frequencies,
\begin{gather}
 \nonumber
 \Delta_1^{(\rm freq)}=(k_p-k_1-k_2)^{(1)}=\\
 \label{freq-correct}
 \frac{A_p(\omega_1+\omega_2-\omega_0)-
 A_h(\omega_1-\omega_h)-A_l(\omega_2-\omega_l)}{c},
\end{gather}
where
\begin{gather}
 \nonumber
 A_h=c\frac{dk_1}{d\omega_1}\Big|_{\omega_1=\omega_h}=\frac{c}{{\rm v}_{gr}^{(p)}},\,
  A_l=c\frac{dk_2}{d\omega_2}\Big|_{\omega_2=\omega_l}=\frac{c}{{\rm v}_{gr}^{(l)}}\\
 \label{Ap-A12}
  A_p=c\frac{dk_p}{d\omega_p}\Big|_{\omega_p=\omega_0}=\frac{c}{{\rm v}_{gr}^{(p)}}.
\end{gather}
In these equations ${\rm v}_{gr}^{(p)}$, ${\rm v}_{gr}^{(h)}$, and ${\rm v}_{gr}^{(p)}$ are group velocities of the pump and of the higher-frequency and lower-frequency emitted photons.

The frequency contributions to the phase mismatch (\ref{freq-correct}) can be present in a slightly different and somewhat more convenient form:
\begin{equation}
 \label{freq-mismatch via A pm}
 \Delta_1^{(\rm freq)}=A_+(\omega_1+\omega_2-\omega_0)-
 A_-(\omega_1-\omega_2- \xi\omega_0)
\end{equation}
with
\begin{equation}
 \label{freq-mismatch via A pm}
 A_+=A_p-\frac{A_h+A_l}{2}\;{\rm and}\;
 A_-=\frac{A_h-A_l}{2}.
\end{equation}
Note that this expansion becomes insufficient in the limit $\xi\rightarrow 0$, i.e., in the frequency-degenerate case. In this limit two emitted photons become identical, their group velocities coincide and $A_-$ of Eq. (\ref{freq-mismatch via A pm}) turns zero. This means that in the frequency-degenerate case the dependence of $\Delta_1^{(\rm freq)}$ on the difference of frequency variables $\omega_1-\omega_2$ disappears, and to find this dependence one has to take into account dispersion, i.e., much smaller second-order corrections to the frequency-dependent mismatch. Such procedure was used in the works \cite{Mikh,Brida}. But in the frequency-nondegenerate regimes considered in this work central frequencies and group velocities of two emitted photons are different from each other, owing to which the dependence of the mismatch on the difference of frequencies is present already in the first-order expansion, and any small second-order corrections are not needed.

\begin{figure}[h]
\centering\includegraphics[width=8 cm]{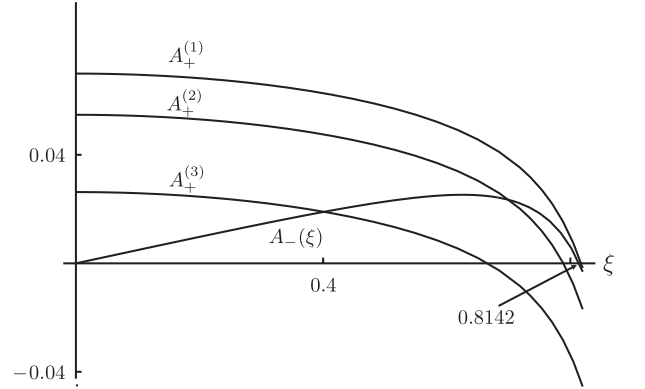}
\caption{{\protect\footnotesize {The functions $A_-(\xi)$ and $A_+(\varphi_0,\xi)$, the latter at $\varphi_0=0.37734\,(1), 0.500578\,(2)$ and $0.7\,(3)$.}}}\label{Fig6}
\end{figure}

The functions $A_+(\xi;\varphi_0)$ and $A_-(\xi)$ are shown in Fig. \ref{Fig6}.
This picture shows in particular that the function $A_-(\xi)$ turns zero not only at $\xi=0$ but also at the point $\xi=0.8142$. As mentioned above this point is very special because this a point where the curve $n_{\rm eff}(\xi)$ has its maximum (Fig. \ref{Figl}) and the curve $\varphi_0^{\rm Coll}(\xi)$ (Fig. \ref{Fig2}) has its minimum.
Significance of vicinity of the point $\xi=0.8142$ for spectral features of emitted photons will be discussed below in Section 6.

\begin{figure}[h]
\centering\includegraphics[width=8 cm]{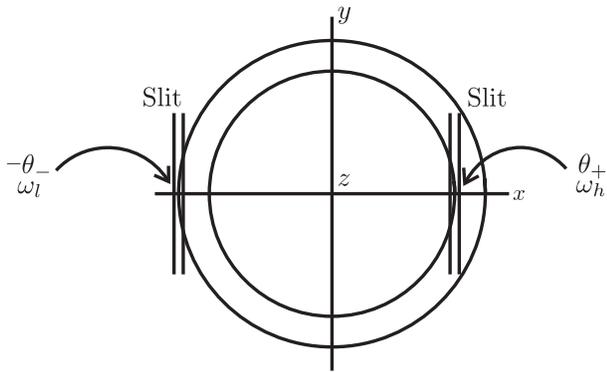}
\caption{{\protect\footnotesize {A two-slit scheme of measurements in a given plane $(x,z)$.}}}\label{Fig7}
\end{figure}

Now, with all derivations done, we can write down explicitly the total angular-frequency biphoton wave function.
\begin{gather}
 \nonumber
 \Psi \propto \exp\left[-\frac{(\omega_1+\omega_2-\omega_0)^2\tau^2}{2}\right]\times\\
 \nonumber
 \exp\left[-\frac{({\widetilde\theta}_{1\,x}
 +{\widetilde\theta}_{2\,x})^2w^2\pi^2}{2\lambda_p^2}\right]
%\times
 %\nonumber
{\rm sinc}
\Bigg\{\frac{{\widetilde\theta}_{1\,x}-{\widetilde\theta}_{2\,x}-2\theta_0}{(\delta\theta)_L}
+\\
\nonumber
\frac{L}{2c}\Big[A_+(\omega_1+\omega_2-\omega_0)-A_-(\omega_1-\omega_2- \xi\omega_0)\Big]\Bigg\}\times\\
\nonumber
F_{sl}(\theta_{1\,x}-\theta_+)F_{sl}(\theta_{2\,x}+\theta_-)\\
\label{W.F. total}
+\Big(1\leftrightarrows 2\Big),
\end{gather}
where $\tau$ is the pump-pulse duration, $F_{sl}$-functions are form-factors of slits, and the expression in the last line means repeating the same what is written in four first lines but with transposed numbers of angular and frequency variables, $\theta_{1,2\,x}\rightarrow\theta_{2,1\,x}$ and $\omega_{1,2}\rightarrow\omega_{2,1}$. Note also, that the wave function of Eq.  (\ref{W.F. total}) corresponds to measurements in a single given plane $(x,z)$ and, for simplicity, for the case of opening only two slits of four shown in the scheme of Fig. \ref{Fig5}. The two-slit scheme of measurement is shown schematically in Fig. \ref{Fig7}. Generalizations for the case a four-slit scheme of Fig. \ref{Fig5}  will be discussed later after a series of simplifications.

Let us assume now that slits have a width $(\delta\theta)_{sl}$ smaller than $(\delta\theta)_L$. Under this condition the terms with angular variables in the argument of the sinc-function are on the order of $(\delta\theta)_{sl}/(\delta\theta)_L\ll 1$, owing to which these terms can be dropped. Moreover, if we are not interested in detailed analysis of narrow angular distributions of photons after slits, we can roughen our description by indicating only directions of photon propagation at positive or negative angles $\theta_x$. Mathematically this means that the product of slit form factors $F_{sl}(\theta_{1\,x}-\theta_+)F_{sl}(\theta_{2\,x}+\theta_-)$ can be replaced by the product of two columns, $\left (1\atop 0\right)_1\left (0\atop 1\right)_2$, with the upper and lower lines corresponding to $\theta_x>0$ and $\theta_x<0$.

At last, let us use also the approximation of long pump pulses, $\tau\gg\tau_{gr}$, where $\tau_{gr}=L|A_+|/2c$ is the characteristic time related to the difference of group velocities of the pump and of the emitted photons. An estimate at $A_+\sim 0.3$ (see the curves for $A_+(\xi)$ in Fig. \ref{Fig6}) gives $\tau_{gr}\sim 0.1\,{\rm ps}$. Hence the pump pulses can be considered as long in the picosecond or longer ranges and the pulses are short in the cases of femtosecond durations. The case of long pump pulses corresponds to small values ($\sim\tau_{gr}/\tau\ll 1$) of the terms $\propto A_+\nu_+$ in the arguments of sinc-functions in the expression of Eq. (\ref{W.F. total})

A simplified in this way two-frequency wave function takes the form
\begin{gather}
 \nonumber
 \Psi(\omega_1,\omega_2)=\Phi(\omega_1,\omega_2)\begin{pmatrix}1\\ 0\end{pmatrix}_1\begin{pmatrix}0\\ 1\end{pmatrix}_2+\\
 \label{W.F. simlif}
 \Phi(\omega_2,\omega_1)\begin{pmatrix}0\\ 1\end{pmatrix}_1\begin{pmatrix}1\\ 0\end{pmatrix}_2,
\end{gather}
where
\begin{gather}
 \nonumber
 \Phi(\omega_1,\omega_2;\xi)\propto\exp\left[-\frac{(\omega_1+\omega_2-\omega_0)^2\tau^2}
 {2}\right]\times\\
 \label{Phi}
{\rm sinc}
\left[\frac{LA_-}{2c}(\omega_1-\omega_2-\xi\omega_0)\right]\,e^{i\omega_1 \Delta t}
\end{gather}
with $\Delta t$ being the delay time of photons moving in the region of positive $\theta_x$. This delay time is introduced here for analyzing in the following section the HOM effect and its peculiarities arising owing to nondegeneracy of the SPDC process.

The two-frequency wave function of Eq. (\ref{W.F. simlif}), (\ref{Phi}) can be used for finding the temporal wave function ${\widetilde\Psi}(t_1,t_2)$ defined as the Fourier transform of $\Psi(\omega_1,\omega_2)$
\begin{equation}
 \label{Fourier}
 {\widetilde\Psi}(t_1,t_2)=\int d\omega_1d\omega_2\Psi(\omega_1,\omega_2)e^{i(\omega_1t_1+\omega_2t_2)},
\end{equation}
where the temporal variables $t_1$ and $t_2$ can be interpreted as the arrival times of emitted photons to the detector or a beamsplitter.
To make integrals in (\ref{Fourier}) calculable analytically, we model the function ${\rm sinc}(x)$ by the Gaussian function $e^{-\alpha x^2}$ with the fitting parameter $\alpha=0.19292$ found from the condition of equal FWHMs. The result of integration can be presented in the form
\begin{gather}
 \nonumber
 {\widetilde\Psi}(t_1,t_2)=N\left\{F(t_1,t_2)\begin{pmatrix} 1\\0\end{pmatrix}_1\begin{pmatrix} 0\\1\end{pmatrix}_2+\right.\\
 \label{Temporal WF}
 \left.F(t_2,t_1)\begin{pmatrix} 0\\1\end{pmatrix}_1 \begin{pmatrix} 1\\0\end{pmatrix}_2 \right\},
\end{gather}
where $N$ is the normalizing factor and the function $F(t_1,t_2)$ is given by
\begin{gather}
 \nonumber
 F^{2\,{\rm slits}}(t_1,t_2)=\exp\left[\frac{i\xi\omega_0}{2}(t_1-t_2+\Delta t)\right]\times\\
 \label{F}
 \exp\left[-\frac{(t_1+t_2+\Delta t)^2}{8\tau^2}-\frac{(t_1-t_2+\Delta t)^2}{4\alpha L^2A_-^2/c^2}\right] .
\end{gather}

Normalization of the wave function ${\widetilde\Psi}(t_1,t_2)$ is determined by the condition $\int dt_1dt_2 {\widetilde\Psi}(t_1,t_2)^\dag{\widetilde\Psi}(t_1,t_2)=1$ which gives
\begin{equation}
 \label{Norm}
 N=\left(2\int dt_1dt_2|F(t_1,t_2)|^2\right)^{-1/2}.
\end{equation}

\section{Hong-Ou-Mandel effect in the case of frequency-nondegenerate noncollineear biphoton states}

As in the usual HOM effect \cite{HOM}, let us assume that photons from the slit with $\theta_x>0$ and $\theta_x<0$ are sent to the 50-50 \% beamsplitter (BS) under the angles $45^\circ$ from opposite sides. Then the beam splitter makes the following transformation of two-column parts of the wave function of Eq. (\ref{Temporal WF}):
\small
\begin{gather}
 \nonumber
  \begin{pmatrix} 1\\ 0 \end{pmatrix}_1\begin{pmatrix} 0\\ 1 \end{pmatrix}_2\Rightarrow
 \frac{1}{2}\left[\begin{pmatrix} 1\\ 0 \end{pmatrix}_1\begin{pmatrix} 1\\ 0 \end{pmatrix}_2-
 \begin{pmatrix} 0\\ 1 \end{pmatrix}_1\begin{pmatrix} 0\\ 1 \end{pmatrix}_2\right]+\\
 \quad\quad\quad\quad\quad\quad\quad +\frac{1}{2}\left[\begin{pmatrix} 1\\ 0 \end{pmatrix}_1\begin{pmatrix} 0\\ 1 \end{pmatrix}_2-
 \begin{pmatrix} 0\\ 1 \end{pmatrix}_1\begin{pmatrix} 1\\ 0 \end{pmatrix}_2\right]
 \label{transform1}
 \end{gather}
\normalsize
and
\small
\begin{gather}
 \nonumber
  \begin{pmatrix} 0\\ 1 \end{pmatrix}_1\begin{pmatrix} 1\\ 0 \end{pmatrix}_2\Rightarrow
 \frac{1}{2}\left[\begin{pmatrix} 1\\ 0 \end{pmatrix}_1\begin{pmatrix} 1\\ 0 \end{pmatrix}_2-
 \begin{pmatrix} 0\\ 1 \end{pmatrix}_1\begin{pmatrix} 0\\ 1 \end{pmatrix}_2\right]-\\
  \quad\quad\quad\quad\quad\quad\quad -\frac{1}{2}\left[\begin{pmatrix} 1\\ 0 \end{pmatrix}_1\begin{pmatrix} 0\\ 1 \end{pmatrix}_2-
 \begin{pmatrix} 0\\ 1 \end{pmatrix}_1\begin{pmatrix} 1\\ 0 \end{pmatrix}_2\right]
 \label{transform2}
 \end{gather}
\normalsize
The first terms after the symbol $``\Rightarrow"$ in Eqs. (\ref{transform1}) and (\ref{transform2}) correspond to unsplit SPDC pairs in which both photons of biphoton pairs propagate together in one of two directions after BS, whereas the terms in the second lines of these equations correspond to split pairs, in which one photon of a pair propagates after BS in one direction and the second one in other direction, orthogonal to the first one (see Fig. \ref{Fig7}). Probabilities of getting unsplit or split pairs after BS are measurable experimentally by counting numbers of photons either in each of two channels separately or in the coincidence scheme in both channels simultaneously. In the ideal HOM effect  the coincidence signal and the probability of splitting turn zero owing to interference, and such cancelation occurs if the incidence photons have identical features (polarization and frequencies) and arrive to BS simultaneously \cite{HOM}. Any deviations from these conditions diminish efficiency of interference and make the coincidence signal different from zero. In the case  of noncollinear and nondegenerate SPDC which we consider here, both frequencies of photons and their arrival times are not strictly given but rather are somewhat uncertain. This is the reason why the analysis of modifications of the  HOM effect for such case is interesting and can give nontrivial results.
\begin{figure}[h]
\centering\includegraphics[width=8 cm]{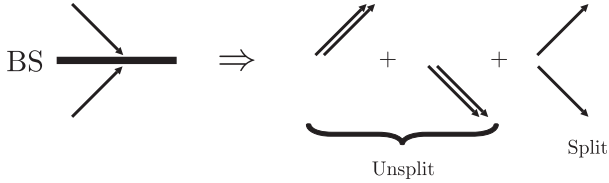}
\caption{{\protect\footnotesize {General possibilities of splitting or propagating unsplit for biphoton pairs after BS.}}}\label{Fig8}
\end{figure}

\begin{figure}[h]
\centering\includegraphics[width=7 cm]{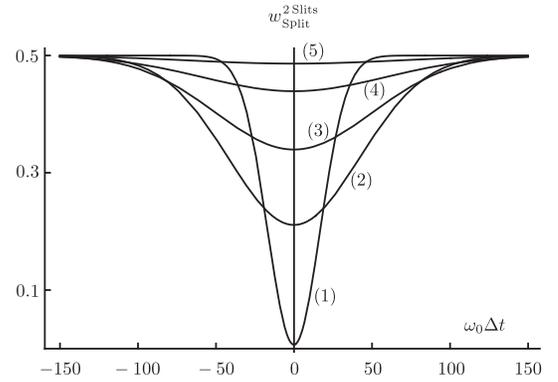}
\caption{{\protect\footnotesize {Probability of splitting photon pairs at the beamsplitter (BS) vs. the delay time $\Delta t$ in the channel $\theta_x>0$ at $\xi=0.01\,(1),\,0.025\,(2), 0.03\,(4), 0.035\,(4)$ and $0.04\,(5)$.}}}\label{Fig9}
\end{figure}

In accordance with the form of the temporal biphoton wave function before the beamsplitter (\ref{Temporal WF}), the probability amplitudes of these processes are given by
\begin{gather}
 A_{\rm unsplit,\,split}(t_1,t_2)=\frac{N}{\sqrt{2}}\Big[F(t_1,t_2)\pm F(t_2,t_1)\Big].
 \label{prob ampl}
\end{gather}
The squared absolute values of these amplitudes determine the differential probability densities of these process
\begin{equation}
 \label{prob dens}
 \frac{dw_{\rm unsplit,\,split}}{dt_1dt_2}=|A_{\rm unsplit,\,split}(t_1,t_2)|^2,
\end{equation}
and the total probabilities are given by integrals over arrival times of both photons
\begin{equation}
 \label{tot prob}
 w_{\rm unsplit,split}=\int dt_1dt_2\frac{dw_{\rm unsplit,split}}{dt_1dt_2}.
\end{equation}
Calculation of integrals is straightforward and the final result for the two-slit scheme of Fig. \ref{Fig7} is given by
\begin{gather}
 \nonumber
 w_{\rm unsplit,split}^{\rm 2\,slits}(\Delta t)=\frac{1}{2}\left\{1\pm\exp\left[-\frac{\xi^2\omega_0^2\alpha L^2A_-^2}{2c^2}\right]\right.\times\\
 \label{2slit}
 \left.\exp\left[-\frac{\Delta t^2}{2\alpha L^2A_-^2/c^2}\right]\right\}.
\end{gather}
Arising in this scheme dependencies of the probabilities of splitting photon pairs after BS, $w_{\rm split}^{\rm 2 slits}(\Delta t)$, are shown in Fig. \ref{Fig9} at a series of values of the nondegeneracy parameter $\xi$.
The curve (1) in Fig. \ref{Fig9} corresponds to a very small degree of nondegeneracy, $\xi=0.01$, and it describes the normal HOM effect: owing to interference, the probablity of getting split pairs almost vanishes at zero delay time $\Delta t=0$ but at longer $|\Delta t|$ it rises up to the level of 0.5. This last case corresponds to accidental reflection from or propagation through the beamsplitter for each photon independently of another one and with no interference. As clearly seen from other curves of Fig. \ref{Fig9}, the increasing  degree of nondegeneracy $\xi$ destroys very quickly the HOM effect in the two-slit scheme of measurements by diminishing the dip of the curves $w_{\rm Split}^{\rm 2\,slits}(\Delta t)$. Already at  $\xi=0.04$ the curve $w_{\rm Split}^{\rm 2\,slits}(\Delta t)$ becomes almost flat, practically without any dip at $\Delta t=0$, which indicates that in this case interference is almost completely missing.

\begin{figure}[t]
\centering\includegraphics[width=8 cm]{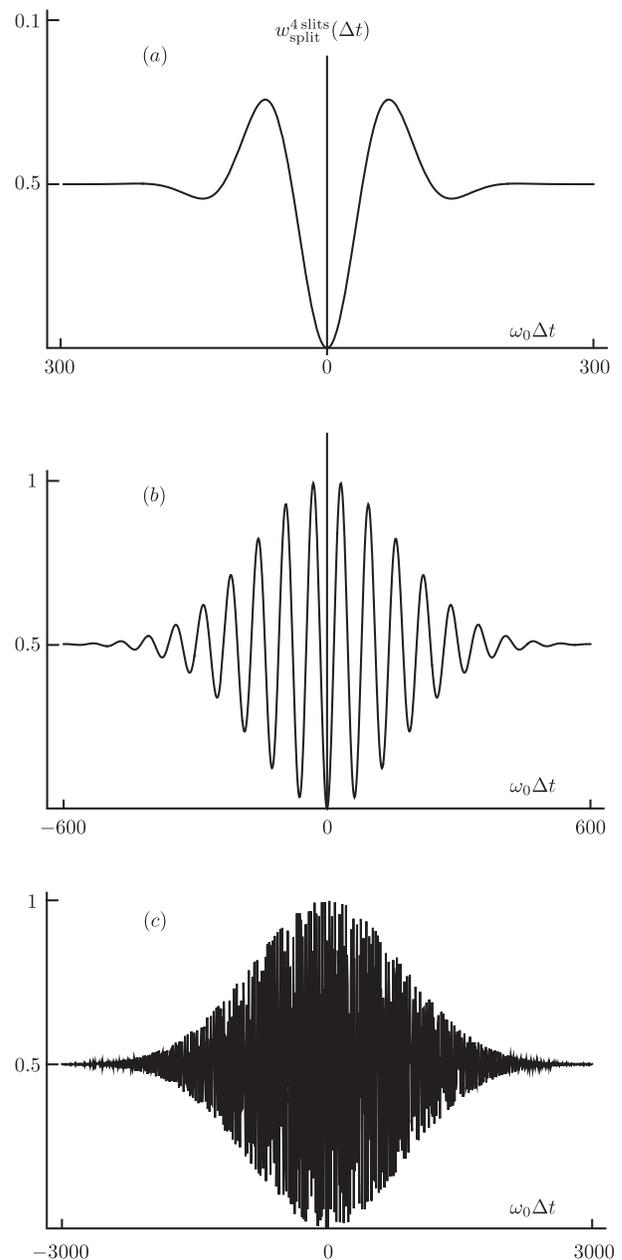}
\caption{{\protect\footnotesize {The probability for biphoton pairs to be split after beamsplitter for a four-slit scheme of measurements in dependence on the delay time $\Delta t$ (in units of $1/\omega_0$) for 5 values of the nodegeneracy parameter $\xi:  0.04\,(a);\, 0.1\, (b); 0.6\, (c)$.}}}\label{Fig10}
\end{figure}

The picture is absolutely different in a scheme of measurements with four slits presented at Fig. \ref{Fig5}.
Mathematically, addition of the second pair of slits corresponding to the same value of the nondegeneracy parameter $\xi$ means the following: if for one pair of slits the wave function is $\Psi(\xi)$, for two pairs of slits it will be equal to $\Psi(\xi)+\Psi(-\xi)$. By applying this rule to the wave function of Eqs. (\ref{Temporal WF}), (\ref{F}), we find that in the case of four slits Eq. (\ref{Temporal WF}) does not change, but in Eq. (\ref{F}) for the function $F(t_1,t_2)$ the exponential factor $\exp\left[\frac{i\xi\omega_0}{2}(t_1-t_2+\Delta t)\right]$ is replaced by $\cos\left[\frac{\xi\omega_0}{2}(t_1-t_2+\Delta t)\right]$,
\begin{gather}
 \nonumber
 F^{\rm 4\,\rm slits}(t_1,t_2)=\cos\left[\frac{\xi\omega_0}{2}(t_1-t_2+\Delta t)\right]\times\\
 \label{F-4}
 \exp\left[-\frac{(t_1+t_2+\Delta t)^2}{8\tau^2}-\frac{(t_1-t_2+\Delta t)^2}{4\alpha L^2A_-^2/c^2}\right] .
\end{gather}
This ``small" change significantly changes final formulas and following from them results. But the general procedure of calculations remains the same as described above in the beginning of this section. So, again, with details of integrations dropped, we find the final result to be given by
\begin{gather}
 \nonumber
 w_{\rm unsplit,split}^{4\,{\rm slits}}(\Delta t)=\frac{1}{2}\left\{1\pm\exp\left[-\frac{\Delta t^2}{2\alpha L^2A_-^2/c^2}\right]\right.\times\\
 \label{4slit}
 \left.
 \frac{\cos(\xi\omega_0\Delta t)+\exp{\left(-\xi^2\omega_0^2\alpha L^2A_-^2/2c^2\right)}}{1+\exp{\left(-\xi^2\omega_0^2\alpha L^2A_-^2/2c^2\right)}}\right\},
\end{gather}
with the correct normalization condition
\noindent $w_{\rm unsplit}^{4\,{\rm slits}}+w_{\rm split}^{4\,{\rm slits}}=1$. In experiment, for finding these probabilities one has to measure the numbers of both split and unsplit pairs, $N_{\rm split}$ and $N_{\rm unsplit}$, and then the probabilities $w_{\rm unsplit,split}$ are defined as
\begin{equation}
 \label{prob via N}
 w_{{\rm split}\atop{\rm unsplit}}=\frac{N_{{\rm split}\atop{\rm unsplit}}}{N_{\rm split}+N_{\rm unsplit}}.
\end{equation}
The most interesting and typical curves of the dependence $w_{\rm split}^{4\,{\rm slits}}(\Delta t)$ (\ref{4slit}) are shown in a series of pictures in Figs. \ref{Fig10} and \ref{Fig12}.
The curve $(a)$ in Fig. \ref{Fig10} is plotted at the same value of the nondegeneracy parameter $\xi=0.04$ at which in the two-slit scheme the HOM effect disappears (Fig. \ref{Fig9}, the curve (5)). The difference between these two curves shows clearly that the addition of photons from the second pair of slits returns interference to the biphoton state under consideration. A deep interference dip at $\Delta t=0$ is present both at $\xi=0.04$ and at higher degrees of nondegeneracy. At all values of the parameter $\xi$ and in all curves of Fig. \ref{Fig10} $ w_{\rm split}^{4\,{\rm slits}}(\Delta t=0)=0$.

Note that there is an evident similarity between the effect of returning interference (at $\Delta t=0$) in the four-slit scheme while it's missing in the scheme with two slits and an analogous effect occurring in a much simpler case of purely polarization states \cite{LPL}. If photons coming to the BS $simultaneously$ from top ($t$) and from bottom ($b$) have identical given frequencies $\omega_0/2$ but different orthogonal polarizations, $H$ (horizontal) and $V$ (vertical), and if the incoming state vector is $a_{H t}^\dag a_{V b}^\dag\ket{0}$, then interference is missing, and the probabilities of getting split and unsplit pairs after BS are equal, $w_{\rm split}=w_{\rm unsplit}=\frac{1}{2}$. If however, the incoming state vector is $\frac{1}{\sqrt{2}}(a_{H t}^\dag a_{V b}^\dag+a_{V t}^\dag a_{H b}^\dag)\ket{0}$, interference returns and this results in $w_{\rm split}=0$ and $w_{\rm unsplit}=1$. Addition of the second term to the state vector in this example is analogous to opening the second pair of slits in the scheme of Fig. \ref{Fig5} compared to the two-slit scheme of Fig. \ref{Fig7}.

A new effect differing the four-slit scheme from the the two-slit one is the appearance of oscillations in the dependencies $ w_{\rm split}^{4\,{\rm slits}}(\Delta t)$ at $\xi\geq  0.04$ and formation of the comb-type structures. If the curve $(a)$ of Fig. \ref{Fig10} (at $\xi=0.04$) can be considered only as a hint for possible existence of the oscillation regime, the curve $(b)$ shows that already at $\xi=0.1$ oscillations are pretty well pronounced. With further growth of the  nondegeneracy parameter $\xi$ the number of oscillations increases as well as the region occupied by them. The curve $(c)$ illustrates the regime of extremely high number of oscillations occurring at $\xi=0.6$.

Note that the curves $(b)$ and $(c)$ of Fig. \ref{Fig10} remind to some extent the curves of Fig. 4 of the work \cite{Abouraddy} though there are big differences both in the problem formulation and in the meaning of curves. In our formulation the pictures of Fig. \ref{Fig10} characterize  the probability of observing split biphoton pairs after BS summed over both photon arrival times $t_1$ and $t_2$. This picture is valid only for the four-slit scheme of measurements and we consider here only the traditional HOM scheme with a single BS and a single varying temporal delay in one of two channels before BS. In our description parameters of the temporal wave function and of the probabilities $w_{\rm unsplit,split}^{4\,{\rm slits}}$ are related to the degree of noncollionearity of SPDC and expressed in terms of the function $A_{\min}(\xi)$ determined by the difference of the photon group velocities in a crystal. In the work \cite{Abouraddy} authors consider a model two-frequency wave function in the frame of the dual-delay scheme with more than one beamsplitter and more than two propagation channels. Besides, the curves in Fig. 4 of \cite{Abouraddy} characterize the expected coincidence signal between photons coming from two different beamsplitters. Resemblance of our results with those of \cite{Abouraddy} occurring in spite of these differences is rather interesting and it emphasizes universality of the underlying interference phenomenon which shows up itself in similar ways at rather pronouncedly different conditions. Moreover, probably it can be said that to some extent the dual-delay scheme with two slits imitates the situation occurring in the scheme with four slits.

In fact, the number of periods of well pronounced oscillations in the curves of Fig. \ref{Fig10} is controlled by relation between their period
 \begin{equation}
 \label{T-osc}
 T_{\rm osc}= \frac{2\pi}{\omega_h-\omega_l}=\frac{2\pi}{\xi\omega_0},
 \end{equation}
 and the time $T_{1/2}$ it takes for the probability $w_{\rm split}^{4\,{\rm slits}}(\Delta t)$ to reach the regions where $w_{\rm split}^{4\,{\rm slits}}\approx 1/2$. The time $T_{1/2}$ is determined by the first exponent on the right-hand side of Eq. (\ref{4slit}),
\begin{equation}
 \label{T1/2}
 T_{1/2}=T_{\rm decoh}=\sqrt{2\alpha}\frac{LA_{\min}(\xi)}{c}=\sqrt{\frac{\alpha}{2}}
 \left(\frac{L}{v_{gr}^{(h)}}-\frac{L}{v_{gr}^{(l)}}\right).
\end{equation}
The time $T_{1/2}-T_{\rm decoh}$,  is on the order of the difference between times required for the higher- and lower-frequency photons to propagate from the beginning to the end of the crystal in which they are produced. This time can be referred to as the decoherence time because at $\Delta t >T_{1/2}$ both probabilities $w_{\rm split}^{4\,{\rm slits}}(\Delta t)$ and $w_{\rm unsplit}^{4\,{\rm slits}}(\Delta t)$ become equal, and equal to 1/2. This is the case when both photons of all biphoton pairs behave as independent particles showing no coherence or interference.

\begin{figure}[h]
\centering\includegraphics[width=7 cm]{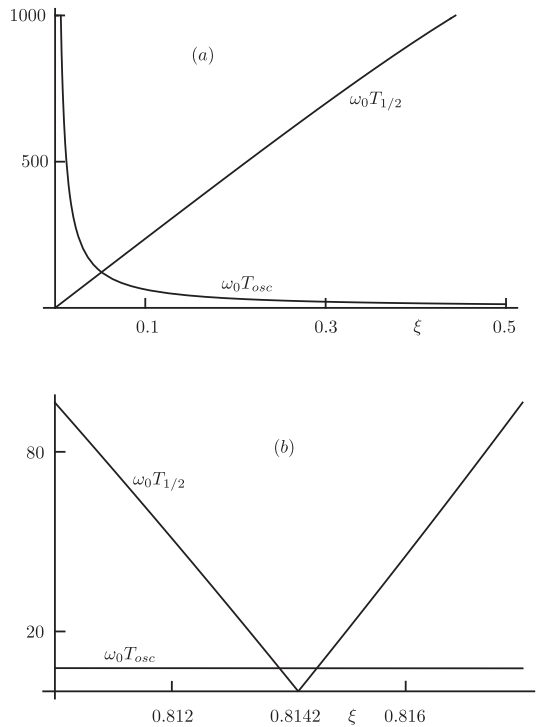}
\caption{{\protect\footnotesize {Oscillation period of the curves in Fig. \ref{Fig10} (\ref{T-osc}) and the time it takes for $w_{\rm split}^{4\,{\rm slits}}(\Delta t)$ to reach the level 1/2 (\ref{T1/2}), both in units of $\omega_0^{-1}$; $(a)$ in the region of small and medium values of the nondegeneracy parameter $\xi$ and $(b)$ in a small region close to $\xi = 0.8142$.  }}}\label{Fig11}
\end{figure}

\begin{figure}[h]
\centering\includegraphics[width=8 cm]{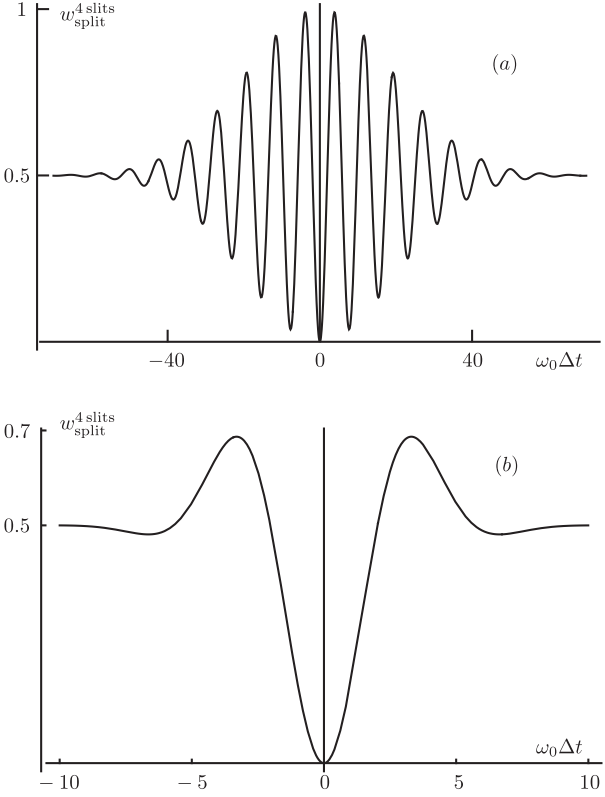}
\caption{{\protect\footnotesize {The function $w_{\rm split}^{4\,{\rm slits}}(\Delta t)$ (\ref{4slit}) at $\xi$ close to 0.8142}}}\label{Fig12}
\end{figure}

Returning to the oscillation regimes of the HOM effect, it's evident that the number of observable oscillations is determined by the ratio of the duration of the oscillation regime  to the period of oscillations, $T_{1/2}/T_{\rm osc}$. Both $T_{1/2}$ and $T_{\rm osc}$ are shown in Fig. \ref{Fig11}$(a)$ (in units of $\omega_0^{-1}$) as functions of the nondegeneracy parameter $\xi$.
It's clear that there are no oscillations in the region of very small values of $\xi$ because in this case $T_{\rm osc}\gg T_{1/2}$, i.e. the period of hypothetically possible oscillations is much longer than the region where they can exist. It's clear also that oscillations start appearing at $T_{\rm osc}\sim T_{1/2}$, which corresponds to $\xi\approx 0.05$ in agreement with the curve $(a)$ of Fig. \ref{Fig10}. At last,  in the region of a higher degree of nondegeneracy ($\xi\geq 0.1$) the ratio $T_{1/2}/T_{\rm osc}$ and the number of observable oscillations are high and are growing with growing $\xi$.

However, an interesting effect occurs when the nondegeneracy parameter $\xi$ is even higher and approaches the point $\xi_0=0.8142$ where the function $A_{\min}(\xi)$ turns zero (see Fig. \ref{Fig6}). Close to this point the decoherence time $T_{1/2}$ becomes very small and can become comparable again with the oscillation period $T_{\rm osc}$, which can be seen in Fig. \ref{Fig11}$(b)$ where the times $T_{1/2}$ and $T_{\rm osc}$ are plotted as functions of $\xi$ in a small vicinity of the point $\xi=0.8142$. Behavior of the probability $ w_{\rm split}^{4\,{\rm slits}}(\Delta t)$ in this region of the nondegeneracy parameter $\xi$ is illustrated by two pictures of Fig. \ref{Fig12}, which look very similar to the first two curves of Fig. \ref{Fig10}.
However, in spite of similarity, there is a rather well provounced difference in scaling of curves in Figs. \ref{Fig10} and \ref{Fig12}. For example, if the distance between two peaks in Fig. \ref{Fig11}$(a)$ equals $\delta(\omega_0\Delta t)=140$, in the similar curve in Fig. \ref{Fig12}$(b)$ $\delta(\omega_0\Delta t)=6.7$, i.e., in fact the curve of Fig. \ref{Fig12}$(b)$ is 20 times narrower than the curve of Fig. \ref{Fig10}$(a)$.

Note, that oscillations in the dependencies on the delay time $\Delta t$ of the coincidence signals after BS have been observed experimentally \cite{Shi-Serg}, mostly for the type-II phase-matching regimes. However, as we know, formation of the finite-size temporal combs of Figs. \ref{Fig10} and \ref{Fig12} has never been seen in experiments. In this context, the main qualitative difference between the type-I and type-II phase-matching regimes concerns the group velocities of the emitted photons. In the type-II case the difference between the ordinary-wave and extraordinary-wave emitted photons is intrinsically present even in the frequency-degenerate regime, owing to which $v_{\rm gr}^{(o)}\neq v_{\rm gr}^{(e)}$ even at $\xi=0$, whereas in the type-I case the difference between the group velocities arises only in the non-degenerate regimes with $\xi\neq 0$. As the difference of group velocities is essentially important for formation of the comb-like structures in the type-I regimes, existence of such or similar results in the type-II regimes requires a special analysis to be done and reported elsewhere.

Note also that, in principle, the general expressions (\ref{2slit}) and (\ref{4slit}) for the probabilities of getting split/unsplit pairs in both two- and four-slit schemes of measurements could be obtained directly from the two-frequency wave function of Eqs. (\ref{W.F. simlif}) and (\ref{Phi}) and from their extension for the four-slit scheme. The temporal wave functions used above provide additional information about dynamics of evolution of biphoton states. In particular, the temporal wave function of Eqs. (\ref{Fourier})-(\ref{F}) can be used to describe the coincidence differe4ntial probability density $dw^{(c)}/d(t_1-t_2)$ in its dependence on the difference of the two photon's arrival times $t_1-t_2$. For the four-slit scheme of measurements the probability density $dw_{4\,\rm slits}^{(c)}/d(t_1-t_2)$ is given by the integrated over $t_1+t_2$ squared difference of the functions $F^{\rm 4\,\rm slits}(t_1,t_2)$ and $F^{\rm 4\,\rm slits}(t_2,t_1)$ (\ref{F-4})
\begin{gather}
 \nonumber
  \frac{dw_{4\,\rm slits}^{(c)}}{d(t_1-t_2)}\propto\\
  \nonumber
  \Bigg |\cos\left[\frac{\xi\omega_0}{2}(t_1-t_2+\Delta t)\right]
  \exp\left[-\frac{(t_1-t_2+\Delta t)^2}{4\alpha L^2A_-^2/c^2}\right]-\\
  \label{4-slit}
  \cos\left[\frac{\xi\omega_0}{2}(t_1-t_2-\Delta t)\right]\exp\left[-\frac{(t_1-t_2-\Delta t)^2}{4\alpha L^2A_-^2/c^2}\right]\Bigg |^2.
\end{gather}

\begin{figure}[h]
\centering\includegraphics[width=8 cm]{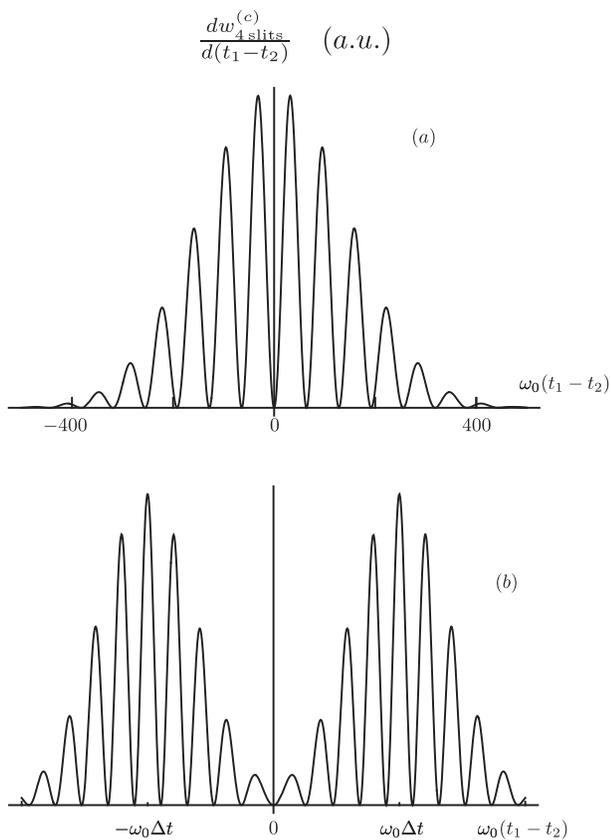}
\caption{{\protect\footnotesize {Four-slit coincidence differential probability $d_{\rm split}^{\rm 4\,slits}/d(t_1-t_2)$ (\ref{4slit}) as a function of $\omega_0(t_1-t_2)$ at $\xi=0.1$; $(a)\,\omega_0\Delta t= 100$ and $(b)\,\omega_0\Delta t=300$ }}}\label{Fig13}
\end{figure}

The dependence of $dw_{4\,\rm slits}^{(c)}/d(t_1-t_2)$ on $t_1-t_2$ is presented in two pictures of at Fig. \ref{Fig13} for two different values of the delay time $\Delta t\neq 0$. These pictures indicate a well pronounce oscillatory structure of the coincidence probability density.
Durations of individual narrow peaks in these curves $\omega_0\delta t$ are on the order of $T_{\rm osc}$ (\ref{T-osc}) determining the period of oscillations in the dependence of the total coincidence probability on the delay time $\Delta t$ (Eq. (\ref{4slit}) and Fig. (\ref{Fig10})). The widths of combs in Fig. \ref{Fig13} are on the order of the decoherence time of Eq. (\ref{T1/2}). Positions of central peaks of the well separated combs in Fig. \ref{Fig13} correspond to $t_1-t_2=\pm\Delta t$. The two combs are well separated if $\Delta t>T_{\rm decoh}$ and they merge into a single comb at $\Delta t<T_{\rm decoh}$. At $\Delta t=0$ Eq. ( \ref{4-slit}) gives immediately $\frac{dw_{4\,\rm slits}^{(c)}}{d(t_1-t_2)}\equiv 0$.

\begin{figure}[h]
\centering\includegraphics[width=8 cm]{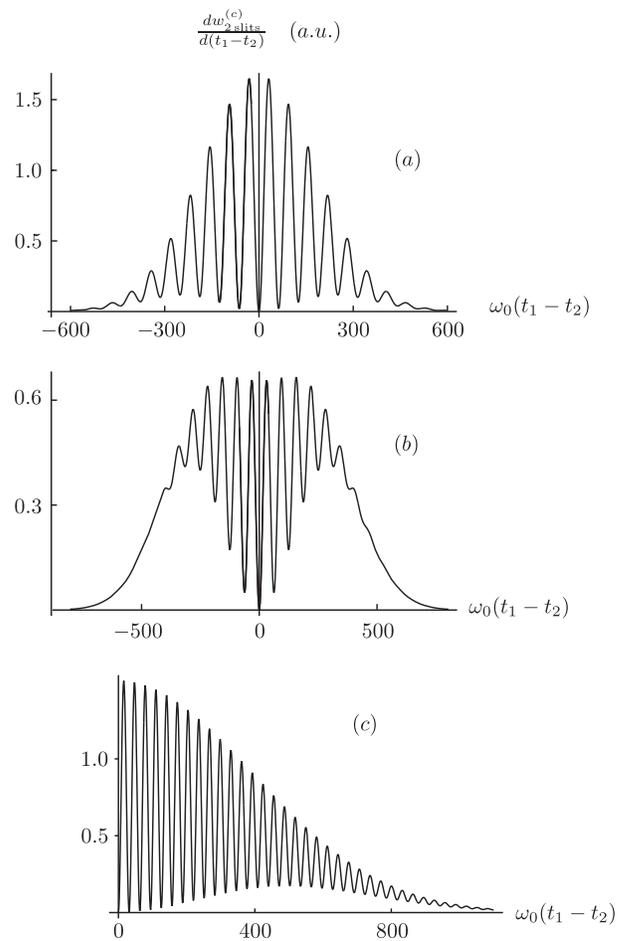}
\caption{{\protect\footnotesize {Two-slit coincidence differential probability
\,
$d_{\rm split}^{\rm 2\,slits}/d(t_1-t_2)$ (\ref{4slit}) as a function of $\omega_0(t_1-t_2)$ at $(a)\,\xi=0.1,\,\omega_0\Delta t= 100$, $(b)\,\xi=0.1,\,\omega_0\Delta t=200$ and $(c)\, \xi=0.2,\,\omega_0\Delta t=200$ }}}\label{Fig14}
\end{figure}

In the case of a two-slit scheme, the coincidence probability density $dw_{2\,\rm slits}^{(c)}/d(t_1-t_2)$ is determined by the  integrated over $t_1+t_2$ squared absolute value of the difference $F^{2\,{\rm slits}}(t_1,t_2)-F^{2\,{\rm slits}}(t_2,t_1)$ with  $F^{2\,{\rm slits}}(t_1,t_2)$ given by Eq. (\ref{F}). The result can be reduced to the form
\begin{gather}
\nonumber
 \frac{dw_{2\,\rm slits}^{(c)}}{d(t_1-t_2)}\propto 2\exp\left[-\frac{(t_1-t_2)^2+\Delta t^2}{2\alpha L^2A_-^2/c^2}\right]\times\\
 \label{ch}
 \left\{\cosh\left[\frac{(t_1-t_2)\Delta t}{\alpha L^2A_-^2/c^2}\right]-\cos\left[\xi\omega_0(t_1-t_2)\right]\right\}.
\end{gather}
The dependence of $dw_{2\,\rm slits}^{(c)}/d(t_1-t_2)$ on $t_1-t_2$ determined by Eq. (\ref{ch})  is shown in Fig. \ref{Fig14} for three different groups of parameters $\xi$ and $\omega_0\Delta t$.
The curves indicate appearance of many oscillations and, generally, their structures are determined by interplay of three characteristic time parameters, the decoherence time $T_{\rm decoh}$ (\ref{T1/2}), period of oscillations $T_{\rm osc}$ (\ref{T-osc}) and the delay time $\Delta t$. The regime with many oscillations forming a single temporal comb in the picture $(a)$ of Fig. \ref{Fig14} keeps almost the same form even at $\Delta t=0$, which differs the case of a two-slit scheme from the four-slit one where $\left.d_{\rm split}^{\rm 4\,slits}/d(t_1-t_2)\right|_{\Delta t=0}\equiv 0$ as clearly seen from Eq. (\ref{4-slit}).

Note that the curves of Fig. \ref{Fig14} remind strongly those of the work \cite{Abouraddy-2}, which were found, however, for absolutely different variables and distributions: for coincidence distributions of photons in transverse coordinates at the crystal exit, i.e., in our notations, for the function $dw^{(c)}/d(x_1-x_2)$ vs. $x_1-x_2$. In principle, such distribution could have nothing in  common with the described above coincidence temporal distribution after transformation at the beamsplitter. But the results look very similar! We believe that this similarity shows that our results summarized in Fig. \ref{Fig14} and the results of Ref. \cite{Abouraddy-2} have the same origin related to coherence intrinsically present in biphoton states and showing up itself in different possible schemes of measurements in the predicted coincidence interference plots with many oscillations.

Note also that seeing experimentally oscillations in  dependencies on $t_1-t_2$ may be a problem for the present-day technologies because this would require not existing now photon counters with femtosecond temporal resolution. If, however, temporal resolution of counters is longer than the period of oscillations $T_{osc}$ but shorter than the delay time $\Delta t$ and the decoherence time $T_{\rm decoh}$ (\ref{T1/2}), one will be able to see in experiments smooth envelopes of the curves in Fig. \ref{Fig13}, described theoretically in our previous work \cite{LPL}.

\section{Conclusion}

As a resume, the following main results were obtained in the frame of the carried out systematic general analysis of the noncollinear nondegenerate SPDC process.

The degrees of nondegeneracy and noncollinearity of SPDC processes were characterized, correspondingly, by the parameters $\xi=\frac{\omega_h-\omega_l}{\omega_0}$ (\ref{ksi}) and $\theta_0$ (\ref{theta-0}). These parameters were found to be related to each other just by Eq. (\ref{theta-0}) combined with the definitions of the effective refractive indices $n_{\rm eff}$ (\ref{n-eff}) and $N_{\rm eff}$ (\ref{N-eff}). At given values of $\xi$ and of the angle $\varphi_0$ between the crystal optical axis and the pump propagation direction $0z$, SPDC emission has a two-cone form with the higher-frequency photons propagating along the inner cone and the lower-frequency ones - along the outer cone, as shown in Fig. \ref{Fig3}. The opening angles of cones are given by $\theta_{\rm inner,\,outer}=\theta_0(\xi, \varphi_0)/(1\pm\xi)$. If the nondegeneracy parameter $\xi$ is not controlled at all, all different emission cones exist together and the SPDC emission has a complex multicolor and multidirectional form. As shown, a much simpler double-cone structure corresponding to some chosen value of the nondegeneracy parameter $\xi$ can be obtained by means of the angular selection. In the simplest case of measurements in any given plane $(x,z)$ containing the pump propagation axis $0z$, the angular selection can be realized by installation of slits as shown in Figs. \ref{Fig5} and \ref{Fig7} (four- and two-slit schemes, correspondingly). For these schemes we found the biphoton wave function depending on two frequencies of photons $\omega_1$ and $\omega_2$, localized in small vicinities around the central frequencies $\omega_h$ and $\omega_l$. Parameters of this wave function depend on the degree of nondegeneracy $\xi$ or, in other words, on location of slits.

Fourier transformation in both frequencies  $\omega_1$ and $\omega_2$ gives the temporal biphoton wave function, arguments of which are two arrival times of photons $t_1$ and $t_2$ to detectors or to a beamsplitter. In schemes of measurements in a given  plane $(x,z)$ the delay time $\Delta t$ is assumed to be introduced in one of two (or two of four) channels for photons coming to the beamsplitter from one of two sides, and in this way the HOM effect is analyzed in details for nondegenerate noncollinear regimes of SPDC. In the two-slit scheme the nondegeneracy is shown to be destroying the HOM effect: the HOM dip at $\Delta t=0$ is shown to disappear rather quickly with a growing degree degree of nondegeneracy $\xi$ (Fig. \ref{Fig9}). In contrast, in the case of the four-slit scheme the HOM dip at $\Delta t=0$ is found to be present at any degrees of nondegeneracy. This means that transition from the two-slit to four-slit schemes returns coherence of biphoton states lost in the two-slit scheme at $\xi\geq 0.04$. Coherence of biphoton states in the four-slit scheme shows itself also in multiple oscillations of the integral probabilities of getting divided photon pairs after BS (Figs. \ref{Fig10} and \ref{Fig12}). The period of oscillations (\ref{T-osc}) is related to the inverse difference of the central frequencies $\omega_h$ and $\omega_l$ and the amount of observable oscillations is controlled by the decoherence time (\ref{T1/2}) determined by the difference of the group velocities of the higher- and lower-frequency photons and, consequently,  by the difference of propagation times of these photons in the crystal.

Oscillations of the same type occur not only in the integral probabilities (integrated over the arrival times $t_1$ and $t_2$) but also in the probability densities of $dw/d(t_1-t_2)$ in their dependence on the difference of the arrival times $t_1-t_2$ (Figs. \ref{Fig13} and \ref{Fig14}). The most interesting result of this part is the formation of finite-size temporal combs filled with multiple oscillations inside.

We believe that the described results reveal rather interesting and important fundamental features of biphoton states related to their coherence and showing up themselves in temporal interference structures presented above.

\section*{Acknowledgement}
The work was supported by the Russian Science Foundation, grant 14-02-01338

%\medskip
\bibliographystyle{unsrtnat}

\bibliography{Text}

\end{document}